# Predictability of reset switching voltages in unipolar resistance switching

S. B. Lee, S. C. Chae, S. H. Chang, and T. W. Noh[a]

*ReCOE and FPRD, Department of Physics and Astronomy, Seoul National University, Seoul 151-747, Republic of Korea*

In unipolar resistance switching of NiO capacitors, Joule heating in the conducting channels should cause a strong nonlinearity in the low resistance state current–voltage (*I–V*) curves. Due to the percolating nature of the conducting channels, the reset current $I_R$, can be scaled to the nonlinear coefficient $B_o$ of the *I–V* curves, i.e., $I_R \propto B_o^{-x}$. This scaling relationship can be used to predict reset voltages, independent of NiO capacitor size; it can also be applied to TiO$_2$ and FeO$_y$ capacitors. Using this relation, we developed an error correction scheme to provide a clear window for separating reset and set voltages in memory operations.

---

[a] Electronic mail: twnoh@snu.ac.kr.

Unipolar resistance switching (RS) shows reversible bistable resistance states when voltage of the same polarity is applied. It has attracted renewed interest due to its potential application in the nonvolatile memory devices known as resistance change memory (RRAM).[1,2] However, to apply unipolar RS to a commercial nonvolatile memory device, several reliability-related scientific and technical issues must be resolved. For RRAM, the major current challenge is reducing the error signals due to fluctuations in the switching parameters.[2] As shown in Fig. 1, the required external voltages for switching between the bistable resistance states fluctuate too much to secure a reliable operational voltage window. Considerable effort has been expended to solve this fluctuation problem,[3–5] but the lack of fundamental understanding regarding the switching mechanism has remained a major obstacle.

Conventional memory devices, such as flash memory and dynamic random access memory, also have error signal problems; the common solution has been to apply error correction techniques with the assistance of advanced circuit designs.[6] To develop a similar error correction scheme in RRAM, we must find a simple and easily-accessible physical phenomena that can predict the error signals reliably.

Here, we report our attempts at finding a simple physical relationship governing the switch from the low (LRS) to high resistance state (HRS) known as the reset process.

Due to Joule heating in the conducting channels, we found that the current–voltage (*I–V*) curves in the LRS deviated from an Ohmic response and became nonlinear. Although the reset currents $I_R$ fluctuated by about two orders of magnitude, they could be scaled to the nonlinear coefficient of the *I–V* curves. This simple relationship allows us to predict the reset switching voltages reliably, allowing development of a practical error correction scheme for unipolar RRAM.

The 232 *I–V* curves of the NiO capacitor represented in Fig. 1(a) show a very wide reset voltage $V_R$ (low-to-high RS) and set voltage $V_S$ (high-to-low RS) distribution. To examine the switching voltage distributions during successive RS operations, we plotted $V_R$ and $V_S$ according to cycle number in Fig. 1(b). $V_R$ and $V_S$ fluctuated widely from 0.3 to 1.4 V and 1.1 to 4.7 V, respectively. The largest value of $V_R$ and the smallest value of $V_S$ were 1.39 V and 1.12 V, respectively. These values indicate the impossibility of making a clear window for the reset and set operation voltages. Overlaps of $V_R$ and $V_S$ will result in irresolvable errors in the RRAM operations.

The inset in Fig. 1(a) shows eight *I–V* curves in the LRS, chosen from the 232 successive RS operations. Note that the $V_R$ values fluctuated between 0.5 and 0.9 V, and that the $I_R$ value showed even larger fluctuations. Each curve shows a linear Ohmic response at low voltage but a deviation from Ohmic response near $V_R$. To perform a more quantitative analysis, we evaluated the resistance $R$ ($\equiv V/I$) of the LRS. As shown in the inset of Fig. 2(a), all eight $I$–$V$ curves in the LRS can be collapsed into a single line by normalizing with proper values of $V_o$, where the nonlinearity starts to develop. The collapse in the inset of Fig. 2(a) indicates that there should be a governing mechanism which determines the nonlinear response of the LRS.[7]

Joule heating has been well established to play a very important role in LRS.[1,8–13] As the percolating conducting filament in LRS is metallic, its local resistance value $r$ should be proportional to its local temperature $T$: $r(T) = r_o(T_1)[1 + \beta(T - T_1)]$, where $\beta$ is the temperature coefficient and $T_1$ is a fixed reference temperature.[9,13] The local current $i$ flowing in part of the filament should provide Joule heating with power of $i^2 r(T)$, raising $T$ accordingly. Then, $(r - r_o) \propto i^2$. In the percolating network, most of its current will be concentrated in a few singly connected bonds,[14] so $R = R_o + B_o I^2$, where $R_o$ is the resistance in the limit of $V = 0$ V.[15] As shown in Fig. 2(a), all of the curves from the eight LRS fit quite well with $R = R_o + B_o I^2$, indicating the importance of Joule heating in the reset process. Note that $B_o$, which can be easily measured from a few experimental points on the $I$–$V$ curve, is a good parameter to represent the Joule heating effects.

Recently, we demonstrated that the connectivity of the conducting filaments in unipolar RS should be similar to those in classical percolation systems.[5,11,12] In two-dimensional (2D) semicontinuous metal films (i.e., a classical percolating system), studies reported that Joule heating could result in an electric breakdown in the percolating network.[14,16] The Joule heating could also cause nonlinear responses, including third-harmonic generation.[14–16] By measuring the third harmonic coefficient $B_{3f}$ (i.e., the ratio between third-harmonic voltages and the cube of the applied ac current), Yagil *et al.* showed that the breakdown current should be scaled with $B_{3f}$ in 2D semicontinuous metal films.[14,16] Note that $B_o$ is the zero-frequency limit of $B_{3f}$.[15] Therefore, we adopted a simple analogy with our experimental data by replacing $B_{3f}$ with $B_o$. As shown in Fig. 2(b), where $\log_{10}(I_R)$ vs. $\log_{10}(B_o)$ is plotted for our NiO capacitors, we found that $I_R \propto B_o^{-x}$ with $x = 0.3 \pm 0.1$ holds throughout the full range of $B_o$.

Furthermore, we found that the simple scaling relationship between $I_R$ and $B_o$ holds well, independent of sample size and material system. As indicated by the solid circles in Fig. 2(b), the transport data from the 30 × 30 $\mu$m$^2$, 500 × 500 nm$^2$, and 100 × 100 nm$^2$ Ti-doped NiO capacitors[17] follow the same scaling behavior. Note that the scaling relationship $I_R \propto B_o^{-x}$ is valid over a wide range of $B_o$ (about 10 orders of magnitude). Furthermore, as shown in the inset of Fig. 2(b), measurements of $I_R$ and $B_o$ data for FeO$_y$ and TiO$_2$ capacitors[5] yielded $x = 0.32 \pm 0.05$ and $0.37 \pm 0.05$, respectively. Random circuit breaker network model-based simulations[5,13] also yielded $x = 0.3 \pm 0.1$ (data not shown). These agreements suggest that the scaling relationship of $I_R \propto B_o^{-x}$ could be universal and therefore applicable to other RRAM material systems. Note that the $I_R \propto B_o^{-x}$ scaling relationship suggests that if we enter the LRS via a set process, we can reliably predict the value of $I_R$ for the next reset process by simply measuring $B_o$.

Based on this predictability, we developed error correction logic for RRAM. Figure 3(a) shows our proposed error correction scheme flowchart. Using the scaling relationship $I_R \propto B_o^{-x}$, the $I_R$ value for the next reset process can be estimated. We can also estimate the resistance value $R_R$ for the next reset process from $R_R = R_o + B_o I_R^2$. Using the relationship $V_R = I_R R_R = I_R(R_o + B_o I_R^2)$, we can predict the $V_R$ value, i.e., $V_R$(Pre). Then, if $V_R$(Pre) $< V_R^{th}$, the LRS is kept. To change the LRS to HRS, reset switching should occur below $V_R^{th}$; however, if $V_R$(Pre) $\geq V_R^{th}$, the LRS is cleared by applying a large enough voltage pulse for the reset and set operations.[8,17] The RRAM device then enters the next LRS and the comparison should be redone.

We applied our proposed error correction logic to the data given in Fig. 1. Figure 3(b) shows a clear voltage window between $V_R$ and $V_S$ with $V_R^{th} = 0.9$ V. $V_R$(Pre) agreed with the experimental $V_R$ values within a relative error of 10%. This removal process guaranteed that all of the LRS values used in the real RRAM operations will be reset with the application of voltages less than 0.9 V.

In summary, we investigated the nonlinear behavior of the current–voltage curves for a Pt/NiO/Pt capacitor, which showed unipolar resistance switching. In the low resistance state, Joule heating effects cause quite strong deviations from Ohmic behavior. Due to the fractal nature of the conducting filaments in unipolar resistance switching, a scaling relationship $I_R \propto B_o^{-x}$ should exist. Based on this simple physical relationship, we proposed a simple error correction scheme that could provide a clear voltage window for reset and set operations in RRAM.

We thank S. Seo for sample preparation. This work was financially supported by the Creative Research Initiatives (Functionally Integrated Oxide Heterostructure) of the Ministry of Science and Technology (MOST) and the Korean Science and Engineering Foundation (KOSEF). S.B.L. acknowledges financial support from a Seoul Science Scholarship.

FIG. 1. (Color online) (a) The 232 $I$–$V$ curves of a NiO capacitor showing unipolar resistance switching. The red and blue lines correspond to the reset and set processes, respectively. To prevent permanent dielectric breakdown during the set processes, we limited the maximum current, called the compliance current $I_{comp}$, to 1 mA. The inset shows eight LRS-selected $I$–$V$ curves. Note that each curve shows a nonlinear response when $V$ approaches the reset voltage $V_R$. $I_o$ and $V_o$ are the current and voltage at which nonlinearity begins to develop, respectively, and $R_o$ is the resistance in the limit of $V = 0$ V. The current value corresponding to $V_R$ is called the reset current $I_R$. (b) Wide distributions of $V_R$ and set voltage $V_S$ during 232 successive reset and set processes. Voltage overlaps of $V_R$ and $V_S$ result in irresolvable error signals.

FIG. 2. (Color online) (a) Plots of $R$ vs. $I^2$. As all the $R$–$I^2$ curves are straight lines, fitting by $R = R_o + B_o I^2$ yields $B_o$ and $R_o$ for each LRS from the slope and the $y$-intercept. The inset shows plots of normalized resistance $R/R_o$ and normalized voltage $V/V_o$ for eight LRS values. Note that all of the $R/R_o$–$V/V_o$ curves collapse into one line. (b) Plot of $\log_{10}(I_R)$ vs. $\log_{10}(B_o)$. Note that a single scaling relationship holds within a full variation of $B_o$. As denoted by the red solid circles, this scaling relationship persists for Ti-doped NiO capacitors, whose data are from Ref. 17. The inset shows that $TiO_2$

(triangles) and FeO$_y$ (squares) capacitors have nearly the same $x$ values as the NiO capacitors.

FIG. 3. (Color online) (a) Flowchart to explain the error correction scheme. (b) A clear voltage window opened between $V_R$ and $V_S$ after error correction with a threshold value of $V_R^{th}$ = 0.9 V.

*Figure 1*

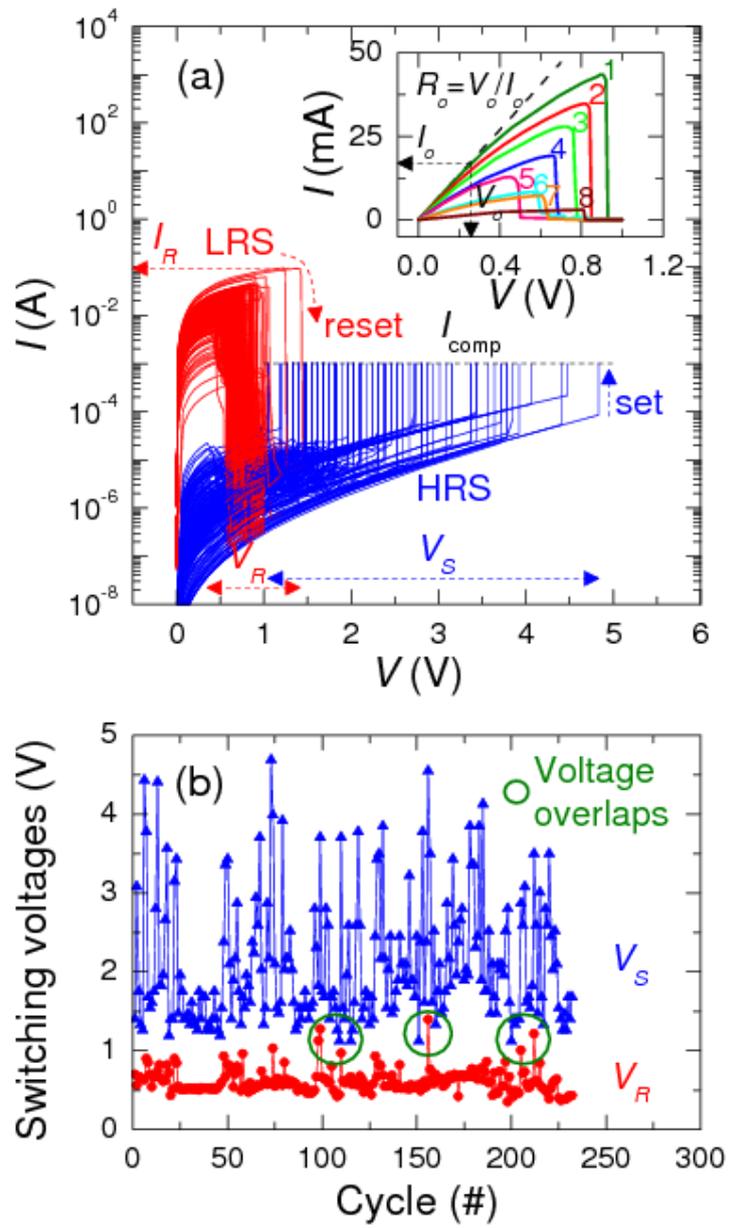

*Figure 2*

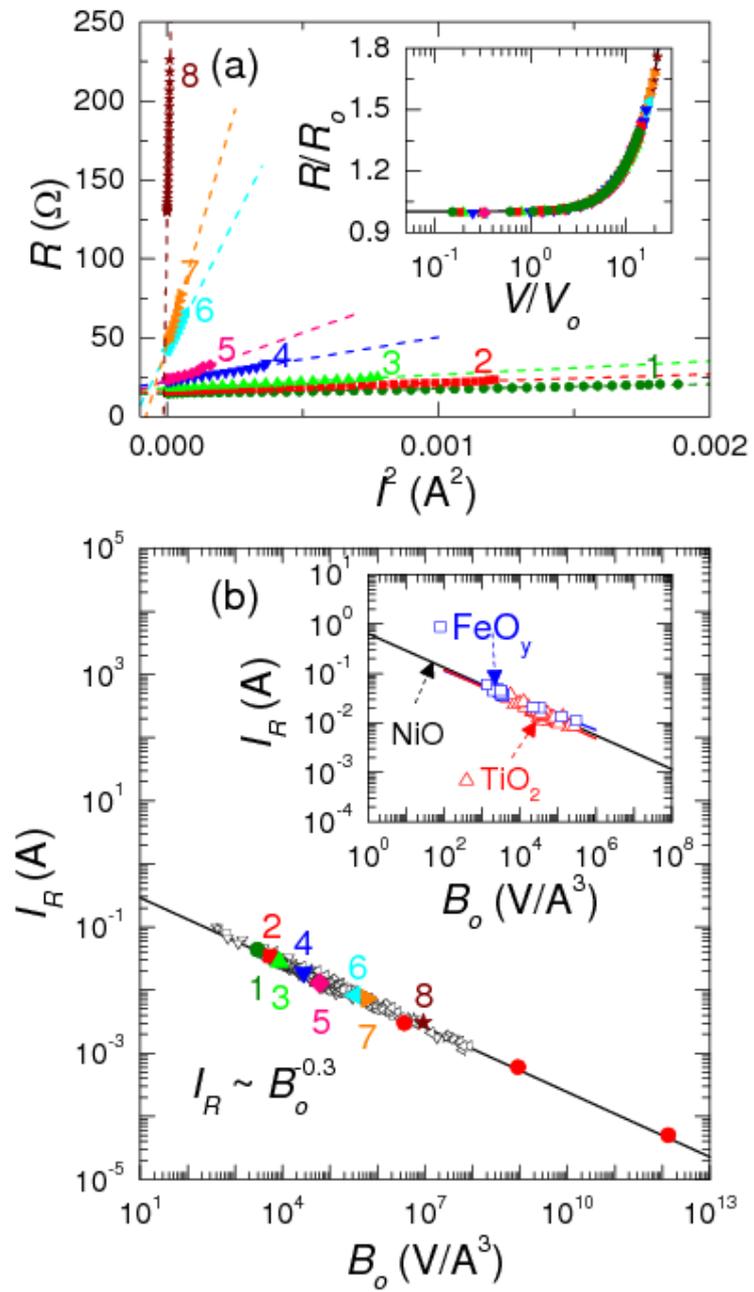

*Figure 3*

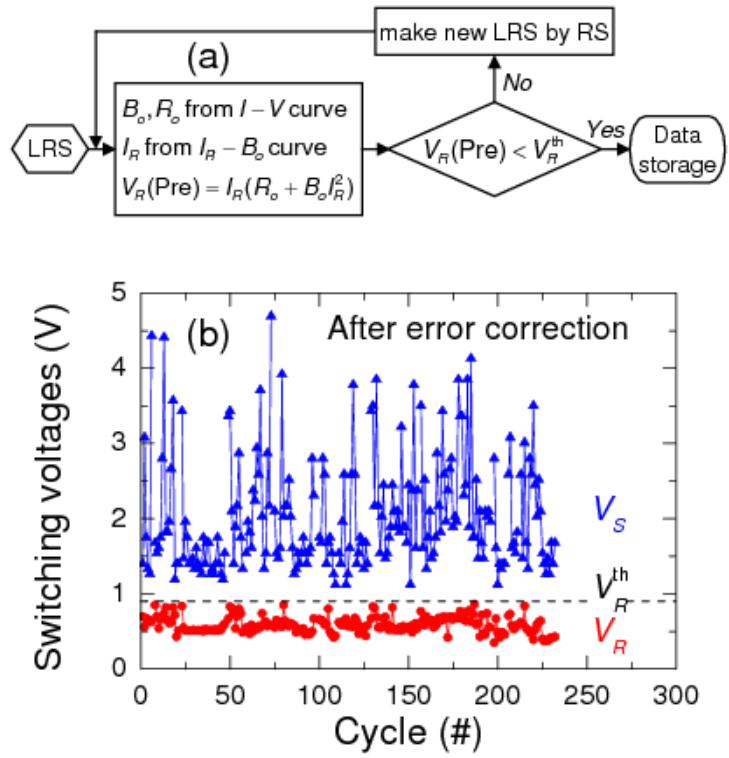